\documentclass[twocolumn,pre,superscriptaddress]{revtex4}

\usepackage{dcolumn}
\usepackage{amsmath}

\usepackage{graphicx}

\newcommand{\half}{\tfrac12}

\newcommand{\defn}{\textit}
\newcommand{\dd}{\mathrm{d}}

\newcommand{\set}[1]{\lbrace#1\rbrace}

\newcommand\cin{c_\textrm{in}}
\newcommand\cout{c_\textrm{out}}

\begin{document}
\title{Efficient method for estimating the number of communities in a network}

\author{Maria A. Riolo}
\affiliation{Center for the Study of Complex Systems, University of
Michigan, Ann Arbor, Michigan, USA}
\author{George T. Cantwell}
\affiliation{Department of Physics, University of Michigan, Ann Arbor,
  Michigan, USA}
\author{Gesine Reinert}
\affiliation{Department of Statistics, University of Oxford, 24--29 St.\,Giles, Oxford, UK}
\author{M. E. J. Newman}
\affiliation{Center for the Study of Complex Systems, University of
Michigan, Ann Arbor, Michigan, USA}
\affiliation{Department of Physics, University of Michigan, Ann Arbor,
  Michigan, USA}

\begin{abstract}
  While there exist a wide range of effective methods for community detection in networks, most of them require one to know in advance how many communities one is looking for.  Here we present a method for estimating the number of communities in a network using a combination of Bayesian inference with a novel prior and an efficient Monte Carlo sampling scheme.  We test the method extensively on both real and computer-generated networks, showing that it performs accurately and consistently, even in cases where groups are widely varying in size or structure.
\end{abstract}

\maketitle

\section{Introduction}
Many networks of interest in the sciences display community structure, meaning that their nodes divide naturally into clusters, modules, or groups, such that there are many network connections within groups but few between groups~\cite{GN02,Fortunato10,FH16}.  The decomposition of networks into their constituent communities is one of the primary tools used for interpreting the structure of large networked systems, allowing us to break data sets apart into manageable pieces and hence make sense of systems that can otherwise defy analysis.

Community detection---the process of identifying good community divisions of a given network---has been the subject of a vigorous research effort in the last 15 years or so, and many different approaches have been proposed.  A fundamental shortcoming of most of them, however, is that they require us to know in advance how many communities a network contains.  We don't usually know this number \textit{a priori}, meaning we need some way to estimate it from the data.  Recently, a number of authors, including ourselves, have proposed methods for making such estimates using Bayesian inference applied to fits of network models to observed network data~\cite{HRT07,DPR08,LBA09,LBA12,MMFH13,CL15,NR16,Yan16,Peixoto17}.  In these approaches, one defines a generative random-graph model of a network with community structure, then fits it to the data to obtain a Bayesian posterior probability distribution over possible divisions of the network into groups, along with the associated number of groups, which we denote~$k$.  Then one averages over this distribution in some way to produce an estimate of the relative probability of different values of~$k$ for the network in question.

In this paper, which builds on our previous work in~\cite{NR16}, we do a number of things.  First, we give a detailed derivation of a practical method for computing the number of groups or communities in real-world network data.  Many of the previous approaches have developed useful formal ideas, but not practical algorithms for real data, because they are based on unrealistic network models---most often the so-called stochastic block model, which is known to be a poor model for calculations of this kind~\cite{KN11a}.  In this paper we employ a more sophisticated model, the degree-corrected stochastic block model, which gives substantially superior results.

Second, we look carefully at the prior probability distribution over divisions of a network into groups, and particularly at generative processes for priors with non-empty groups.  As in many Bayesian approaches, the choice of prior turns out to be crucial to performing useful inference, and in particular we point out that various types of uniform (maximum-entropy) priors, including ones used in previous work, give poor results and should be avoided.  We propose a new prior based on a queueing-type process that appears to give excellent results in our tests.

Third, we describe an efficient Monte Carlo algorithm that exploits specific features of our proposed prior to perform rapid calculations on large networks.

Finally, we give the results of extensive tests of our method on both real and synthetic benchmark networks, which indicate that the method is able to consistently recover known values for the number of communities under real-world conditions.

\section{Degree-corrected stochastic block model}
\label{sec:dcsbm}
The method we propose for estimating the number of communities in a network is based on techniques of statistical inference, in which observed network data are fitted to a generative model of network structure.  The parameters of the fit tell us about the aggregate properties of the data in much the same way that the fit of a straight line through a set of data points can tell us about their slope.

The network model we employ in our calculations is the \defn{degree-corrected stochastic block model}~\cite{KN11a}.  The traditional (non-degree-corrected) stochastic block model, first proposed in the 1980s~\cite{HLL83}, is a simple model for networks with community structure that has been widely studied in the statistics, sociology, and physics literature.  Recently, however, it has been recognized that this model has serious shortcomings~\cite{KN11a} because the networks it generates have a Poisson degree distribution within each community, making them very unlike most empirically observed networks, which typically have highly right-skewed degree distributions.  In practice, this means that the model is often unable to fit observed network data well for any choice of parameter values.  The degree-corrected stochastic block model remedies this problem by introducing additional mechanisms that allow for arbitrary, non-Poisson degree distributions and is found in practice to give much better fits to real-world network data.

In the degree-correct stochastic block model $n$ nodes are divided into some number~$k$ of groups, labeled $1\ldots k$, with $g_i$ denoting the group to which node~$i$ is assigned.  Then edges are placed between nodes with probabilities that depend on group membership.  There are several variants of the model in use that employ slightly different strategies for placing edges, but the most common strategy, and the one we use here, is to place between any node pair~$i,j$ a number~$a_{ij}$ of edges that is Poisson distributed with mean $\theta_i \theta_j \omega_{g_ig_j}$, where $\set{\theta_i}$ and~$\set{\omega_{rs}}$ are sets of parameters whose values we choose.  The numbers of edges~$a_{ij}$ form the elements of the adjacency matrix~$A$ of the network, the standard mathematical representation of network structure.  Although these numbers are Poisson distributed, the expected degrees of the nodes can follow any distribution---within each group they are proportional to the values of the parameters~$\theta_i$, which we are at liberty to choose in any way we please.

The use of a Poisson distribution for the numbers of edges means that the network generated can in theory have multiedges, i.e.,~pairs of nodes connected by more than one edge, which is not usually realistic---most real-world networks do not have multiedges.  Most networks of scientific interest, however, are very sparse, meaning that the values of $\theta_i \theta_j \omega_{g_ig_j}$ are very small and the probability of having two edges between the same pair of nodes is smaller still.  Multiedges are, as a result, few enough in number that they can usually be neglected.  One also normally allows self-edges in the network (edges that connect a node to itself), placing at each node~$i$ a Poisson distributed number of self-edges with mean $\half \theta_i^2 \omega_{g_ig_i}$.  Again this is not realistic but in practice the number of self-edges is small, so they can be neglected.  (By convention, the number of self-edges at node~$i$ is denoted $\half a_{ii}$ and not~$a_{ii}$, i.e.,~$a_{ii}$ is twice the number of self-edges.  The factor of $\half$ in the number of self-edges~$\half \theta_i^2 \omega_{g_ig_i}$ is included for consistency with this definition---it makes the expected value of $a_{ij}$ equal to $\theta_i\theta_j\omega_{g_ig_j}$ for all $i,j$.)

The description above does not completely specify the degree-corrected block model because there remains an arbitrary normalization of the parameters~$\theta_i$ that has yet to be fixed.  We can increase the values of all the $\theta_i$ in group~$r$ by any factor we please and, provided we simultaneously decrease $\omega_{rs}$ by the same factor, the value of $\theta_i \theta_j \omega_{g_ig_j}$, and hence also the model itself, will not change.  We can fix the values of the parameters by choosing a specific normalization for the~$\theta_i$.  A number of choices are possible, all of which are ultimately equivalent, but for our purposes here a convenient choice is to fix the mean of the $\theta_i$ to be~1 in each group:
\begin{equation}
{1\over n_r} \sum_{i=1}^n \theta_i \delta_{r,g_i} = 1,
\label{eq:normalization}
\end{equation}
where $\delta_{ij}$ is the Kronecker delta and $n_r = \sum_i \delta_{r,g_i}$ is the number of nodes in group~$r$.

For the case of given number of group~$k$ and group assignments~$g$ this completes the specification of the model.  With the model specified we can now write down the probability that any particular network with adjacency matrix~$A = \set{ a_{ij} }$ is generated:
\begin{align}
P(A|\omega,\theta,g,k) &= \prod_{i<j} \bigl( \theta_i\theta_j\omega_{g_ig_j} \bigr)^{a_{ij}} e^{-\theta_i\theta_j\omega_{g_ig_j}} 
    \nonumber\\
   &\qquad{}\times
     \prod_i \bigl( \half\theta_i^2\omega_{g_ig_i}
     \bigr)^{a_{ii}/2} e^{-\theta_i^2\omega_{g_ig_i}/2} \nonumber\\
  &\hspace{-6em}{} = \prod_i \theta_i^{d_i} 
  \prod_{r<s} \omega_{rs}^{m_{rs}} e^{-n_r n_s\omega_{rs}}
  \prod_r \omega_{rr}^{m_{rr}} e^{-n_r^2\omega_{rr}/2},
\label{eq:likelihood}
\end{align}
where we have made use of Eq.~\eqref{eq:normalization} in the second equality, $d_i = \sum_j a_{ij}$ is the observed degree of node~$i$, and
\begin{equation}
m_{rs} = \left\lbrace\begin{array}{ll}
         \sum_{ij} a_{ij} \delta_{g_i,r} \delta_{g_j,s} &
         \qquad\mbox{when $r\ne s$,} \\
         \rule{0pt}{13pt}\half \sum_{ij} a_{ij} \delta_{g_i,r} \delta_{g_j,r} &
         \qquad\mbox{when $r=s$,}
         \end{array}\right.
\end{equation}
is the number of edges running between groups $r$ and~$s$.  We have also discarded an overall multiplying constant in Eq.~\eqref{eq:likelihood}, which has no effect on our results.

The parameters~$\theta$ and~$\omega$ are irrelevant to the questions we are interested in and can be integrated out.  To do this we need to fix the prior probabilities on the parameters~$\theta$ and~$\omega$.  We assume the priors to be independent (conditioned on $g,k$), so that $P(\omega,\theta|g,k) = P(\omega|k) P(\theta|g,k)$, and
\begin{equation}
P(A|g,k) = \iint P(A|\omega,\theta,g,k) P(\theta|g,k) P(\omega|k) \>
           \dd\theta \>\dd\omega.
\label{eq:pagk1}
\end{equation}
We employ maximum-entropy (i.e.,~least informative) priors on both $\theta$ and $\omega$.  For $\theta$ this means a uniform prior over the regular simplex of values specified by Eq.~\eqref{eq:normalization}.  For $\omega$ the situation is more complex.  We note that the expected number of edges between groups $r$ and $s$ is
\begin{align}
\sum_{ij} \theta_i \theta_j \omega_{g_ig_j} \delta_{r,g_i} \delta_{s,g_j}
  &= \omega_{rs} \sum_i \theta_i \delta_{r,g_i} \sum_j \theta_j \delta_{s,g_j}
     \nonumber\\
  &= \omega_{rs} n_r n_s,
\end{align}
where we have made use of~\eqref{eq:normalization} again.  But the total number of places one can place an edge between groups $r$ and~$s$ is $n_r n_s$, which means that the average probability of an edge is simply~$\omega_{rs}$.

As mentioned above, most of the networks we look at in practice are very sparse---the average probability of an edge is much less than one.  For this reason a uniform prior on~$\omega_{rs}$ is not appropriate.  Rather, we need a prior that favors values of $\omega_{rs}$ in the vicinity of the average probability of an edge in the network as a whole, which is $p = 2m/n^2$, where $m$ is the total number of edges in the network.  Here, as previously~\cite{NR16}, we use a maximum-entropy prior conditioned on fixing the expected value of $\omega_{rs}$ to be equal to~$p$, which yields the exponential distribution
$P(\omega) = (1/p) e^{-\omega/p}$.

With these choices of priors on $\theta$ and $\omega$, the integrals in Eq.~\eqref{eq:pagk1} can be completed and we get
\begin{align}
P(A|g,k) &= \prod_r n_r^{\kappa_r} {(n_r-1)!\over(n_r+\kappa_r-1)!}
   \nonumber\\
  &{}\times \prod_{r<s} {m_{rs}!\over(pn_r n_s+1)^{m_{rs}+1}}
     \prod_r {m_{rr}!\over (\half p n_r^2+1)^{m_{rr}+1}},
\label{eq:pagk2}
\end{align}
where
\begin{equation}
\kappa_r = \sum_i d_i \delta_{r,g_i}
\end{equation}
is the sum of the degrees of the nodes in group~$r$, and we have again discarded an overall multiplying constant.

\section{Prior on  group assignments}
\label{sec:prior}
Our goal is to use this model as the basis for a Bayesian model selection procedure to estimate the correct value of $k$ for a given network.  To do this, we need to specify a prior on the group assignments, meaning a joint probability distribution $P(g,k)$ on the group labels and the number of groups.  This then allows us to write
\begin{equation}
P(g,k|A) = {P(g,k)P(A|g,k)\over P(A)},
\label{eq:pgka}
\end{equation}
where $P(A|g,k)$ is given by Eq.~\eqref{eq:pagk2}.  Given this distribution, we can either sum over~$k$ to get the posterior distribution on~$g$, which allows us to do community detection, or sum over~$g$ to get the posterior distribution on~$k$, which allows us to choose a value for~$k$.  It is the latter computation that is our primary focus in this paper.  In practice, we cannot perform the sum over~$g$ exactly, but we can approximate it using Markov chain Monte Carlo.

As is often the case with Bayesian methods, the tricky part of the calculation (or one of the tricky parts) is choosing the prior~$P(g,k)$.  It turns out that in the present case this choice can have a substantial impact on the results, making the difference between a method that works well in most cases and a method that does not.  Moreover, some obvious choices of prior, including ones in common use elsewhere in the literature, result in methods that work poorly, so it is not a matter of simply grabbing a well-understood prior off the shelf.

\subsection{Dirichlet prior}
\label{sec:dirichlet}
Suppose for the moment that the number of groups~$k$ is known and let us then ask what the prior $P(g|k)$ on group assignments should be.  (This is not a very realistic assumption, since our whole purpose here is to estimate~$k$, but the exercise is nonetheless instructive as we will see.)  Our first guess at~$P(g|k)$ might be to choose a flat probability distribution: all assignments~$g$ are equally likely, or equivalently every node is equally likely to be in each of the $k$ groups.  This approach is used, for example, in~\cite{GS09}, but, as pointed out by Peixoto~\cite{Peixoto17}, it is unlikely to give good results.  If nodes are distributed with equal probability among the groups then when $n$ is large (as it usually is) the sizes of the groups will be sharply peaked around~$n/k$.  Any state with group sizes significantly different from $n/k$ will occur rarely.  This is strongly at odds with the observed situation in real-world networks, where we commonly encounter heterogeneous group sizes.

An alternative approach, therefore, and the one that is most commonly adopted in the literature, is to assume a uniform distribution not over assignments~$g$ but over the sizes of the groups~$n_r$.  That is, all possible sets of sizes are equally likely, subject only to the constraint that they sum to the size of the whole network: $\sum_r n_r = n$.  The standard way to achieve this is to specify the expected fraction~$\gamma_r\in[0,1]$ of the network occupied by each group~$r$, such that $\sum_r \gamma_r = 1$, then assign nodes to groups independently at random, with probability~$\gamma_r$ of being assigned to group~$r$.  If the $\gamma_r$ are themselves drawn from a uniform distribution, this makes the distribution of the sizes of the groups uniform.  (It is obvious that this makes the \emph{expected} sizes of the groups uniform, since the expected sizes are proportional to~$\gamma_r$, but it is not entirely obvious that the sizes themselves are also uniform.  However, the results given in Section~\ref{sec:nonpara} provide a proof that they are.)

The probability of generating a particular group assignment~$g$ using this process is
\begin{equation}
P(g|\gamma,k) = \prod_{i=1}^n \gamma_{g_i} = \prod_{r=1}^k \gamma_r^{n_r}.
\label{eq:pggammak}
\end{equation}
Since~$\sum_r \gamma_r = 1$, the points defined by the values $\gamma_r$ fall on a regular $(k-1)$-dimensional simplex, which has volume~$1/(k-1)!$.  So a uniform distribution over values of~$\gamma$ has probability density~$P(\gamma|k) = (k-1)!$ and integrating~\eqref{eq:pggammak} over the simplex then gives~\cite{MMFH13,CL15,NR16}
\begin{equation}
P(g|k) = \int P(g|\gamma,k) P(\gamma|k) \>\dd\gamma
       = {(k-1)!\over(n+k-1)!} \prod_{r=1}^k n_r!
\label{eq:dirichlet}
\end{equation}
The uniform distribution over values of~$\gamma$ is a special case of the so-called \defn{Dirichlet prior}, a general prior on a simplex that includes this choice but also includes a spectrum of non-uniform choices as well.

\subsection{Non-parametric prior}
\label{sec:nonpara}
The derivation in the previous section is a standard one, but in a sense it is needlessly complicated.  If our goal is simply to choose group sizes such that all choices are equally likely then assign nodes to those groups at random, why not just do so directly, without introducing other parameters?  There are ${n+k-1\choose k-1}$ possible choices of $k$ groups such that their sizes sum to~$n$.  Let us choose uniformly among these, then the number of ways of placing the nodes in the groups is given by the multinomial coefficient $n!/\prod_r n_r!$, so the probability of any given assignment~$g$ of nodes to groups is
\begin{equation}
P(g|k) = {1\over{n+k-1\choose k-1}\,n!/\prod_r n_r!}
       = {(k-1)!\over(n+k-1)!} \prod_{r=1}^k n_r!
\label{eq:nonpara}
\end{equation}
which recovers Eq.~\eqref{eq:dirichlet} without the need for the parameters~$\gamma_r$.

(A corollary of this result is that the Dirichlet process of Section~\ref{sec:dirichlet} does indeed generate a uniform distribution over possible group sizes, as claimed, since the distributions~\eqref{eq:dirichlet} and~\eqref{eq:nonpara} are identical.)

\subsection{Non-empty groups}
\label{sec:nonempty}
A possible objection to these methods for generating assignments~$g$ (widely used though they are) is that they allow groups to be empty.  It is unclear what the meaning is of an empty group.  If someone were to hand you a network with two clear groups in it, then tell you that really there are three groups but one of them is empty, you might justifiably say that this is not a meaningful statement.

One advantage of the non-parametric formulation of Section~\ref{sec:nonpara} is that generalizes easily to the case where groups are required to be non-empty.  One need simply replace the binomial coefficient for the number of ways of generating the group sizes with the corresponding coefficient for non-empty groups, which is ${n-1\choose k-1}$.  Then
\begin{equation}
P(g|k) = {1\over {n-1\choose k-1}\,n!/\prod_r n_r!}.
\label{eq:nonempty}
\end{equation}

\subsection{Choice of the number of groups}
\label{sec:queueing}
We turn now to the choice of prior~$P(k)$ on the number of groups itself.  Again one's first guess at a prior might be a flat distribution with all choices equally likely.  For non-empty groups as in Section~\ref{sec:nonempty}, the possible choices for number of groups range from 1 to~$n$, so a flat prior would have $P(k) = 1/n$ in this range and zero for all other values of~$k$.  This choice has been made in some previous work~\cite{CL15}, but we find it to give poor results, placing too much weight on high values of~$k$ and significantly overestimating the number of groups in well-understood test cases.  In practice one must use a strongly decreasing prior on~$k$ to achieve consistent results.  Previous authors have given qualitative arguments in favor of a prior going as~$1/k!$~\cite{MMFH13} or even steeper~\cite{Peixoto14a}.

In this paper we take a somewhat different approach and do away with an explicit prior on~$k$, instead employing a generative process for group assignments~$g$ that automatically incorporates the choice of the number of groups in a simple way.  The process we use, a queueing-type mechanism which is a variant on the ``restaurant'' processes of traditional probability theory, is as follows.  Take the $n$ nodes in random order and place the first one in group~1.  Then for each subsequent node either (a) with probability $1-q$ place it in the same group as the previous node or (b) with probability~$q$ make it the first node in the next group.  Note that this process never generates an empty group.  All groups contain at least one node.

The number of possible orders of the nodes in this process is~$n!$, with each one occurring with equal probability~$1/n!$.  If the process generates~$k$ groups in total then there must be $k-1$ new groups started and, since every node except the first has equal chance $q$ of starting a new group, the probability of generating $k$ groups with sizes $n_1\ldots n_k$ is
\begin{equation}
(1-q)^{n_1-1}q(1-q)^{n_2-1}q\ldots q(1-q)^{n_k-1}
  = q^{k-1} (1-q)^{n-k},
\end{equation}
where we have made use of the fact that $\sum_r n_r = n$.  Furthermore, there are $\prod_r n_r!$ ways of rearranging the nodes within each group that give rise to the same assignment~$g$.  Hence the probability of generating any given assignment under our proposed process is
\begin{equation}
P(g,k) = {1\over n!}\,q^{k-1} (1-q)^{n-k} \prod_{r=1}^k n_r!
\label{eq:pgk0}
\end{equation}
Given that $P(g,k) = P(k) P(g|k)$ and comparing with Eq.~\eqref{eq:nonempty}, we see that this process is equivalent to the process of Section~\ref{sec:nonempty} if one chooses a prior~$P(k)$ on the number of groups thus:
\begin{equation}
P(k) = {P(g,k)\over P(g|k)} = {n-1\choose k-1} q^{k-1} (1-q)^{n-k},
\label{eq:pk}
\end{equation}
with $1\le k\le n$.  In other words, the number of new groups $k-1$ created in the generating process has a binomial distribution (as one can easily derive by considering the process directly).

For our purposes it will be convenient to parametrize the probability~$q$ by $q = \mu/(n-1)$, so that $\mu$ is the expected number of new groups started during the assignment process, which is one less than the total number of groups.  Then
\begin{equation}
P(g,k) = {(1-q)^n\over q n!} {\mu^k\over(n-\mu-1)^k} \prod_{r=1}^k n_r!
\end{equation}
The leading factor of $(1-q)^n/qn!$ is independent of both $g$ and $k$ and will cancel out of subsequent calculations.

\subsection{Choice of parameter value}
\label{sec:parameter}
Our prior on $g,k$ now has just one parameter~$\mu$, whose value we have yet to choose.  One way to proceed is to take the Bayesian approach a step further and place a prior on the prior---a so-called hyperprior, meaning a probability distribution on~$\mu$.  In principle one could go even further and place a hyperhyperprior on the hyper\-prior too, and so forth \emph{ad infinitum}.  This process usually yields diminishing returns, however, and one normally stops at some point, simply fixing a value for the parameters.  In the present case, we choose to halt the process at the level of the parameter~$\mu$.  In our tests we have found that a value $\mu=1$ works well, so that, neglecting constants
\begin{equation}
P(g,k) = (n-2)^{-k} \prod_{r=1}^k n_r!
\label{eq:pgk}
\end{equation}
although other values around~1 give basically the same results, so the method does not seem sensitive to the precise choice we make.  No doubt a hyperprior centered roughly around~1, such as a suitably sized normal distribution, would also give similar results, but we don't see any advantage to taking this approach.

It is interesting to note that when $q = \mu/(n-1)$ with $\mu=1$, and taking the limit of large~$n$, the prior on~$k$, Eq.~\eqref{eq:pk}, becomes
\begin{equation}
P(k) = {e^{-1}\over(k-1)!}.
\end{equation}
In other words, this choice is essentially equivalent to the $1/k!$ prior proposed previously on heuristic grounds~\cite{MMFH13}.

\section{Monte Carlo algorithm}
Given the prior, Eq.~\eqref{eq:pgk}, on $g,k$, we can now write down the complete posterior distribution~\eqref{eq:pgka} on the same quantities.  Then by summing over all values of $g$ we can find the probability distribution~$P(k|A)$ and hence deduce the most likely value of~$k$.  Unfortunately the sum over~$g$ is hard to do: it has $k^n$ terms, which is a very large number in most cases, making exhaustive numerical evaluation impossible, and no simple scheme presents itself for performing the sum analytically.  Instead therefore we estimate the distribution over~$k$ by Markov chain Monte Carlo sampling.

The Monte Carlo scheme we propose employs steps of two types:
\begin{itemize}
\item \textbf{Type 1:} Moving a single node from group to group.  This type of move includes processes that decrease the number of groups (if the node moved is the last of its group) and processes that keep the number of groups constant (if the node moved is not the last of its group).
\item \textbf{Type 2:} Moving a single node to a newly created group, thereby increasing the value of~$k$ by one.
\end{itemize}

A sufficient condition for a correct Monte Carlo algorithm is that the algorithm satisfy the requirements of \defn{ergodicity} and \defn{detailed balance}~\cite{NB99}.  The requirement of ergodicity says that every state of the system must be accessible from every other by a finite sequence of Monte Carlo steps.  It is trivial to show that this condition can be satisfied by steps of the kind described above that move individual nodes from one group to another.

More demanding is the requirement of detailed balance, which in the present situation says that the rate $R(g,k\to g',k')$ to go from a state~$(g,k)$ to another state~$(g',k')$ and the rate $R(g',k'\to g,k)$ to go back again must satisfy
\begin{equation}
{R(g,k\to g',k')\over R(g',k'\to g,k)}
   = {P(g',k'|A)\over P(g,k|A)}
   = {P(g',k')\over P(g,k)} \times {P(A|g',k')\over P(A|g,k)},
\label{eq:ratio}
\end{equation}
where we have used Eq.~\eqref{eq:pgka}.  From Eq.~\eqref{eq:pgk} we have
\begin{equation}
{P(g',k')\over P(g,k)} = (n-2)^{k-k'}
  {\prod_{r=1}^{k'} n_r'!\over\prod_{r=1}^k n_r!},
\label{eq:priorratio}
\end{equation}
where $n_r'$ are the sizes of the groups for group assignment~$g'$.

We use a traditional accept/reject Monte Carlo scheme in which we repeatedly propose a potential move then either accept or reject that move with probabilities chosen to satisfy the detailed balance condition.  Thus the rate $R(g,k\to g',k')$ divides into the product of the probability~$\pi$ of proposing the move in question and the probability~$\alpha$ of accepting it: $R(g,k\to g',k') = \pi(g,k\to g',k')\,\alpha(g,k\to g',k')$.  Then
\begin{equation}
{R(g,k\to g',k')\over R(g',k'\to g,k)}
   = {\pi(g,k\to g',k')\over\pi(g',k'\to g,k)} \times
     {\alpha(g,k\to g',k')\over\alpha(g',k'\to g,k)}.
\label{eq:propacc}
\end{equation}

The algorithm we propose is as follows:
\begin{enumerate}
\item \begin{enumerate}
\item On each step of the algorithm, with probability $1-1/(n-1)$ we propose a move of type~1.  Specifically, if $k=1$ we do nothing (because there are no possible moves that move a node from one group to another).  Otherwise, when $k>1$, we choose a pair of distinct group labels~$r,s$ uniformly at random from the set of all such pairs in the range $1\ldots k$, then choose a single node uniformly at random from group~$r$ and move it to group~$s$.
\item If, in the process, we remove the last remaining node from group~$r$, leaving that group empty, we relabel the non-empty groups so that their labels run from $1\ldots k-1$, and we decrease $k$ by one.  In practice, the most efficient way to do the relabeling is just to change the current group~$k$ to have label~$r$ (unless $r=k$, in which case no relabeling is necessary).
\end{enumerate}
\item \begin{enumerate}
\item Otherwise, with probability $1/(n-1)$ we propose a move of type~2.  Specifically, we choose a pair of distinct group labels~$r,s$ uniformly at random from the set of all such pairs in the range $1\ldots k+1$, relabel group~$r$ as group~$k+1$, and create a new empty group~$r$ (unless $r=k+1$, in which case we simply create a new empty group~$k+1$ and no relabeling is necessary).  Then we choose a node uniformly at random from group~$s$ and move it to the newly created group~$r$.
\item If, in the process, we remove the last remaining node from group~$s$, leaving that group empty, we change group~$k+1$ to have label~$s$.  Otherwise we increase $k$ by~1.  (Note that $k$ can never become greater than~$n$ during this process, since doing so would always involve removing the last node from group~$s$, which precludes increasing~$k$ any further.)
\end{enumerate}
\item Once we have proposed our move from state $(g,k)$ to state~$(g',k')$, we accept it with acceptance probability
\begin{equation}
\alpha(g,k\to g',k') = \min\biggl( 1, {P(A|g',k')\over P(A|g,k)} \biggr).
\label{eq:mh}
\end{equation}
If the move is accepted, $g',k'$~becomes the new state of the system.  Otherwise the system remains in the old state~$g,k$.  Note, crucially, that the probabilities appearing on the right-hand side of~\eqref{eq:mh} are of the form $P(A|g,k)$, and not $P(g,k|A)$ as you might expect if you are familiar with standard (Metropolis--Hastings) Monte Carlo methods.
\item Repeat from step~1.
\end{enumerate}

To see that that this algorithm does indeed satisfy the condition of detailed balance, Eq.~\eqref{eq:ratio}, consider first a move of type~1 that moves a node from group~$r$ to group~$s$, where $n_r>1$ so that the node moved is not the last node in group~$r$ and the value of $k$ does not change.  Thus $k'=k$ and Eq.~\eqref{eq:priorratio} simplifies to
\begin{equation}
{P(g',k')\over P(g,k)} = \prod_{t=1}^k {n_t'!\over n_t!} = {n_s'\over n_r},
\label{eq:type1}
\end{equation}
with the terms for all groups other than $r$ and $s$ canceling.  Moves of type~1 are performed with probability $1-1/(n-1)$, there are $k(k-1)$ possible choices of distinct groups~$r,s$, and $n_r$ nodes to choose from in group~$r$, so the total probability of proposing a specific move of a specific node~is
\begin{align}
\pi(g,k\to g',k') &= \biggl( 1 - {1\over n-1} \biggr) \times {1\over k(k-1)}
                     \times {1\over n_r} \nonumber\\
  &= {n-2\over(n-1)k(k-1) n_r}.
\label{eq:type1prop}
\end{align}
Similarly, the probability of proposing the reverse move, in which the same node is moved from group $s$ to group~$r$,~is
\begin{equation}
\pi(g',k'\to g,k) = {n-2\over(n-1)k(k-1) n_s'}.
\end{equation}
And the ratio of the two is
\begin{align}
{\pi(g,k\to g',k')\over\pi(g',k'\to g,k)}
  &= {n-2\over(n-1)k(k-1) n_r} \times {(n-1)k(k-1) n_s'\over n-2} \nonumber\\
  &= {n_s'\over n_r},
\end{align}
which is precisely equal to Eq.~\eqref{eq:type1}.

We can also demonstrate an equivalent result for moves that change the value of~$k$.  Consider a move of type~1 that removes the last node from group~$r$ and moves it to group~$s$, thereby reducing the number of groups by~1, so that $k'=k-1$.  For such a move, Eq.~\eqref{eq:priorratio} becomes
\begin{equation}
{P(g',k')\over P(g,k)} = (n-2)n_s'.
\label{eq:type2}
\end{equation}
The probability of proposing such a move is again given by Eq.~\eqref{eq:type1prop}, except that in this case $n_r=1$, so the expression simplifies to
\begin{equation}
\pi(g,k\to g',k') = {n-2\over(n-1)k(k-1)}.
\label{eq:decreasek}
\end{equation}
The reverse of such a move is a move of type~2, in which a node in group~$s$ becomes the founding member of new group~$r$.  Moves of type~2 are performed with probability~$1/(n-1)$, there are $(k'+1)k' = k(k-1)$ ways of choosing the labels~$r,s$, and $n_s'$ ways of choosing the node to be moved.  Thus the total probability of proposing such a move is
\begin{equation}
\pi(g',k'\to g,k) = {1\over(n-1)k(k-1) n_s'}.
\label{eq:increasek}
\end{equation}
Taking the ratio of~\eqref{eq:decreasek} and~\eqref{eq:increasek}, we get
\begin{align}
{\pi(g,k\to g',k')\over\pi(g',k'\to g,k)}
  &= {n-2\over(n-1)k(k-1)} \times (n-1)k(k-1) n_s' \nonumber\\
  &= (n-2) n_s',
\end{align}
which is equal to~\eqref{eq:type2}.

Finally, there is one further class of moves that have to be considered separately, namely moves of type~2 that remove the last node from group~$s$ and make it the initial node in a new group~$r$.  By contrast with other moves of type~2, these moves do not change the value of~$k$.  Moreover they don't change the product~$\prod_r n_r!$ either, so Eq.~\eqref{eq:priorratio} is simply
\begin{equation}
{P(g',k')\over P(g,k)} = 1.
\label{eq:type3}
\end{equation}
At the same time, the probability of proposing such a move is the same in both directions, equal to $1/[ (n-1)k(k+1) ]$, and hence the ratio of the forward and backward proposal probabilities is~1, which is equal to~\eqref{eq:type3}.

Thus for all moves of all types we have
\begin{equation}
{\pi(g,k\to g',k')\over\pi(g',k'\to g,k)}
  = {P(g',k')\over P(g,k)}.
\label{eq:cancel}
\end{equation}
Equating Eqs.~\eqref{eq:ratio} and~\eqref{eq:propacc} and making use of~\eqref{eq:cancel}, we then find that the detailed balance condition becomes
\begin{equation}
{\alpha(g,k\to g',k')\over\alpha(g',k'\to g,k)}
  = {P(A|g',k')\over P(A|g,k)}.
\label{eq:alpharatio}
\end{equation}
If we can choose the acceptance ratios to satisfy this relation, then Eq.~\eqref{eq:ratio} will be obeyed.  But the choice in Eq.~\eqref{eq:mh} trivially satisfies~\eqref{eq:alpharatio}, hence detailed balance is obeyed and the proposed algorithm will sample correctly from the distribution~$P(g,k|A)$.

\subsection{Implementation}
\label{sec:implementation}
Implementation is a relatively straightforward translation of the algorithm described above into computer code.  We maintain not only a record of the group assignment~$g_i$ of each node but also a separate, unordered list of the members of each group, which allows us to choose a random member of a group rapidly, and to efficiently relabel all members of a group when required.  We calculate the logarithm of the ratio $P(A|g',k')/P(A|g,k)$, rather than the ratio itself, to avoid numerical problems with powers and factorials (which can become large) and only take the exponential at the end to determine the acceptance ratio, Eq.~\eqref{eq:mh}.  We also employ a look-up table of log-factorials to speed their calculation, and a running record of the $n_r$ and~$m_{rs}$, updated after every accepted move, so as to avoid recalculating these values repeatedly.

As a practical matter, the relabeling process for moves of type~2 can be slow if group~$r$ contains many nodes, so in our implementation we always give new groups label~$k+1$, which frees us from having to relabel any nodes other than the one node that is placed in the newly created group.  Technically, this means that our Monte Carlo algorithm does not sample labelings~$g$ with the true probability of Eq.~\eqref{eq:pgk}: group~$k$ will typically be the newest group and therefore smaller on average than the others.  However, the algorithm does still sample \emph{divisions} of the network into groups and values of~$k$ with the correct probability, and these are the only physical quantities we care about.  The group labels themselves have no meaning---they exist only as a mathematical convenience for the purposes of notation.  The only meaningful quantities are the value of $k$ and the division into groups, and these are correctly generated.  (And if one were concerned to sample labelings~$g$ correctly, one could do so easily by taking the labelings generated by the algorithm and randomly permuting the labels.)

The initial assignment of nodes to groups is drawn at random from the prior distribution defined in Section~\ref{sec:queueing}.  In order to avoid any bias in the results, instead of just setting $\mu=1$ we use a value of $\mu$ that is itself chosen randomly.  In the example calculations presented in Section~\ref{sec:applications}, the value of $\mu$ for the initial assignment is chosen uniformly in the interval from 0 to~100, meaning in practice the initial number of groups lies approximately uniformly in this range.  (In the actual Monte Carlo calculation, however, we always use $\mu=1$, as described in Section~\ref{sec:parameter}.)

Our code is implemented in~C, and performs about a million Monte Carlo steps per second on a typical desktop computer (circa 2017), which puts the analysis of large networks, up to hundreds of thousands of nodes or more, within reach.  Our code is available for download on the web---see Ref.~\cite{code}.

\subsection{Relation to previous approaches}
In a previous paper~\cite{NR16} we proposed a slightly different algorithm for determining the number of communities, based on the same principles used here but different in detail.  We expect the present algorithm to be more efficient than the earlier one, primarily because of the way it incorporates the prior on group assignments~$P(g,k)$ into the proposal probability $\pi$ rather than the acceptance probability~$\alpha$.  However, we also believe the derivation given here is more appropriate than that given in the previous paper, particularly in its focus on the prior~$P(g,k)$ on group assignments.

The argument given in the previous paper differs from that given here in two ways.  First, in the previous paper we advocated using a flat prior~$P(k)$ on the number of groups, whereas in this paper we argue for a decreasing prior going as~$1/k!$.  On the other hand, the probability~$P(g|k)$ given in the previous paper omits a factor of $k!$, equal to the number of ways the labels on a given partition of the network can be permuted without changing the partition.  These two factors of $k!$ cancel, leaving the equations essentially unchanged.  Thus the formulas and algorithm given in the previous paper are essentially equivalent to those given here, but the motivation differs, with the arguments given here being, in our opinion, the correct ones.

\section{Example applications}
\label{sec:applications}
In this section we apply our method to a wide range of networks and find it to give good results in most cases.  To evaluate performance under controlled conditions, we test the method on several large sets of computer-generated networks, created using both the stochastic block model and the widely used LFR benchmark.  To test the method under real-world conditions we have also applied it to a range of observed networks, including a number of staples of the community detection canon, as well as a large example network with over $300\,000$ nodes.

\subsection{Computer-generated networks}
In order to explore the performance of our method systematically we have tested it on a range of computer-generate (``synthetic'') networks.  These are networks with known community structure planted within them, generated using random graph models.  Using synthetic networks allows us to vary the number of planted communities and quantify the extent to which our algorithm is able to correctly recover that number.

\begin{figure*}
\begin{center}
\hfill\includegraphics[width=6.5cm]{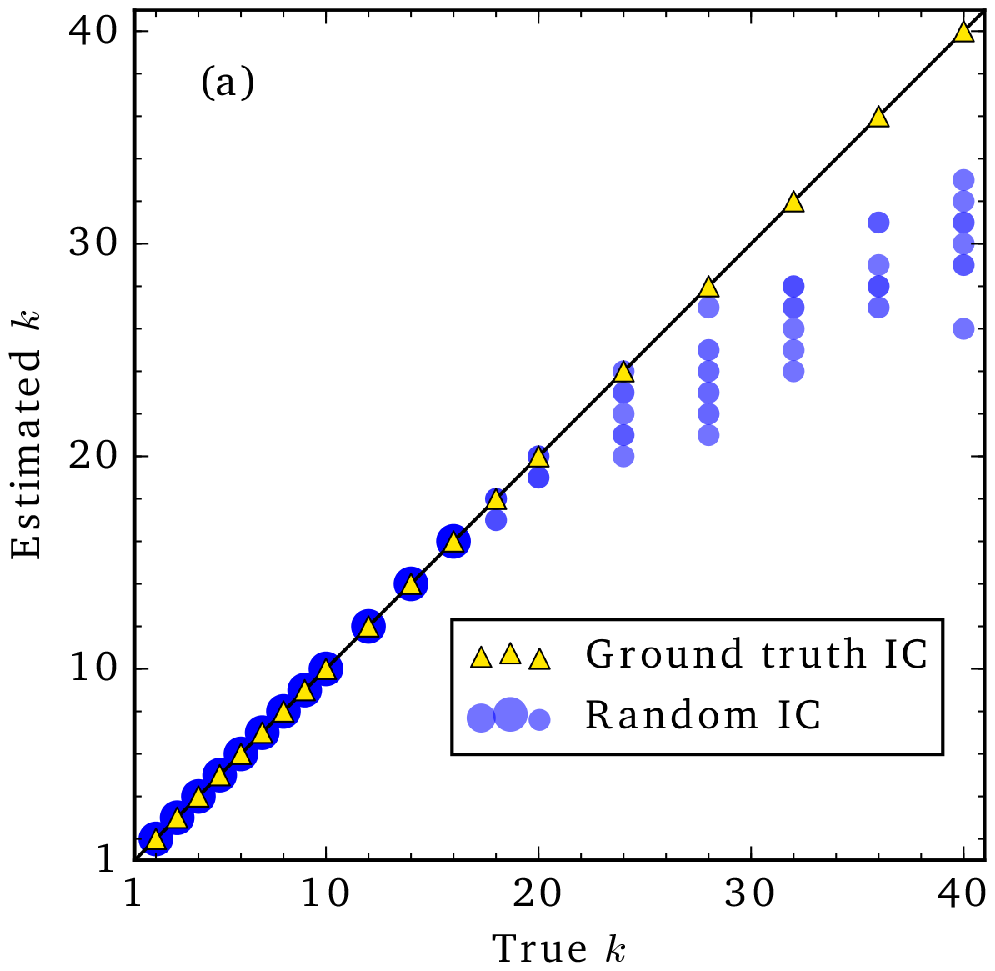}\hfill
\includegraphics[width=6.5cm]{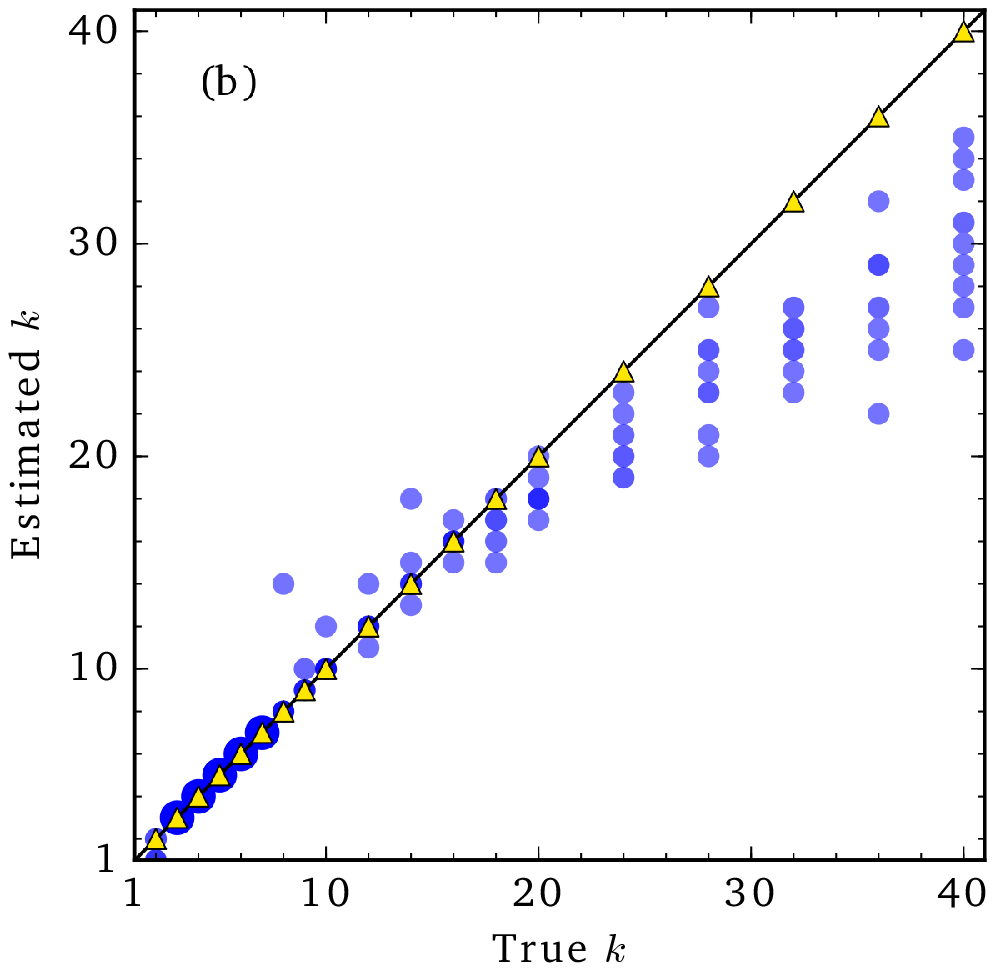}\hfill\null\\
\ \\
\ \\
\hfill\includegraphics[width=6.5cm]{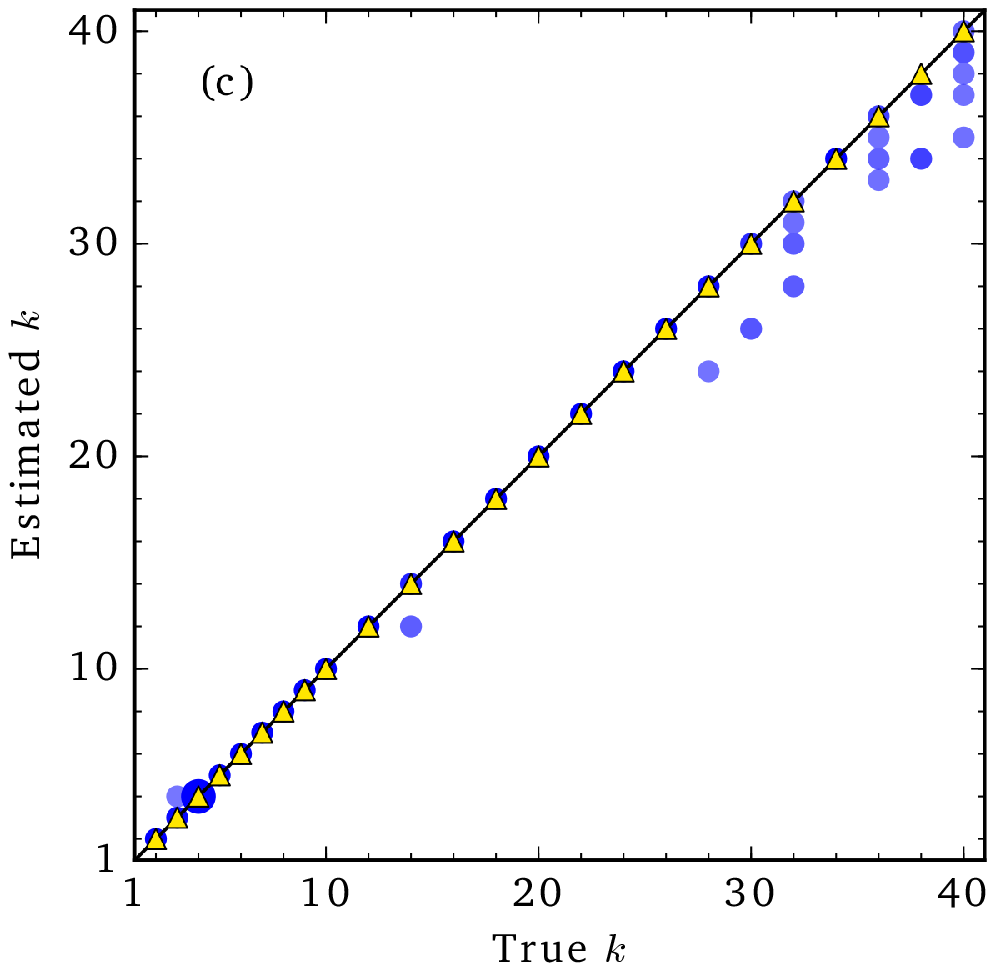}\hfill
\includegraphics[width=6.5cm]{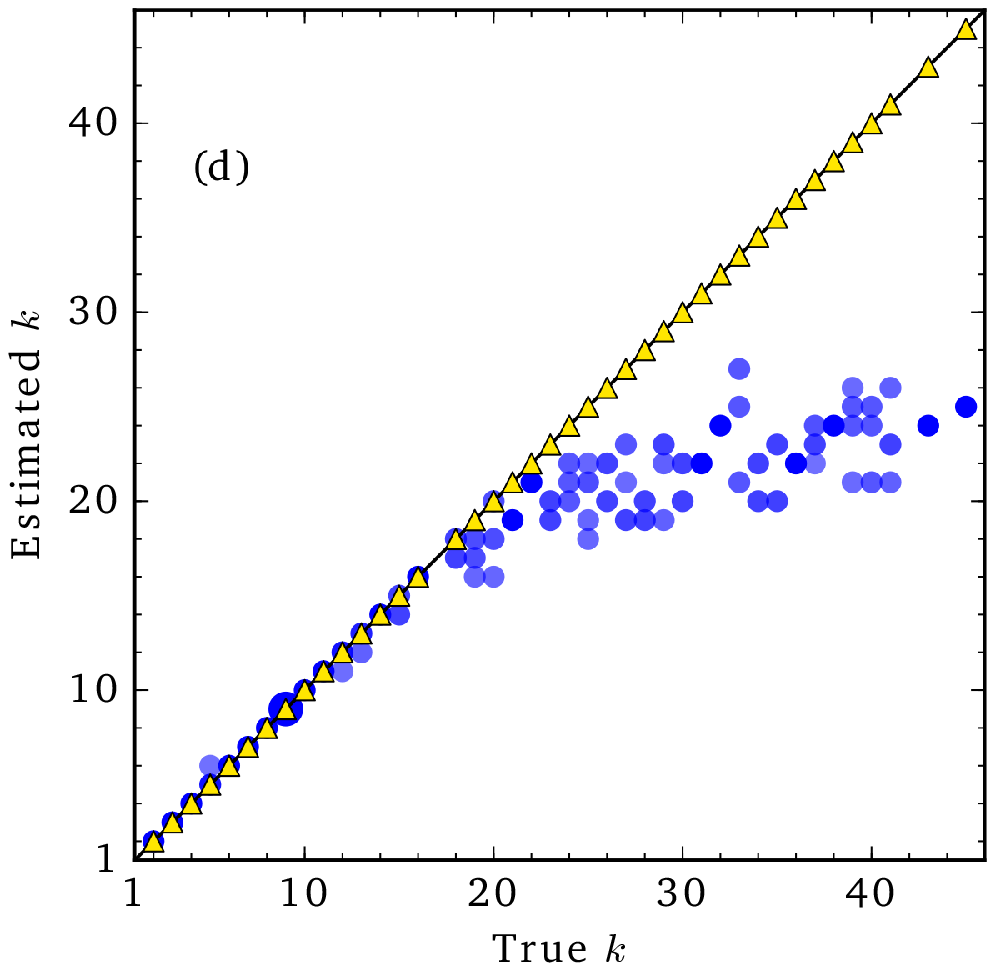}\hfill\null\\
\end{center}
\caption{Tests of the method on synthetic networks.  In each panel, circles represent results derived from Monte Carlo runs with random initial assignment of nodes to group (``random initial conditions''), while triangles represent runs started with the known correct assignments (``ground truth initial conditions'').  (a)~Networks generated using the stochastic block model with $n=1000$ nodes, mean degree~30, and equally sized groups with 90\% of connections with groups and 10\% between groups.  (b)~Networks generated using the stochastic block model with groups of fixed size 250 nodes (so that network size varies with the number of groups~$k$), and each node having an average of 16 in-group connections and 8 out-group connections.  (c)~Networks generated using the same stochastic block model as in (a) but with the rows and columns of the matrix of edge probabilities permuted to produce a mixed assortative/disassortative structure.  (d)~Networks generated using the LFR benchmark model of~\cite{LFR08}, which is parametrized by it maximum and minimum group sizes.  In these tests we used minimum groups sizes between 10 and 80 nodes, and maximum group size equal to five times the minimum.  Ten Monte Carlo runs were performed for each network of 2000 steps per node each, with the distribution over~$k$ being calculated from the final 1000 only.}
\label{fig:synthetic}
\end{figure*}

In our first set of tests, we use networks generated using the standard (non-degree-corrected) stochastic block model~\cite{HLL83,KN11a}.  Figures~\ref{fig:synthetic}a and~\ref{fig:synthetic}b show results for two different sets of networks.  Each panel shows the number of communities~$k$ inferred by our method plotted against the known number planted in the network as the latter is varied (blue circles in the plots).  In Fig.~\ref{fig:synthetic}a the size of the network is held fixed as the number of communities in increased, while in Fig.~\ref{fig:synthetic}b the size of the communities is held fixed, so that the size of the network increases with the number of communities.  As the figures show, in both cases the algorithm infers the correct number of communities with high accuracy for values of $k$ up to about~20.

For higher values it has a tendency to underestimate the number of communities.  These calculations, however, are for runs of the Monte Carlo algorithm that start with a random assignment of nodes to groups, as described in Section~\ref{sec:implementation}.  Also shown in the figures are the results of runs on the same networks in which the Monte Carlo algorithm was started with group assignments corresponding exactly to the planted ground-truth community division (red triangles).  As the figures show, for this choice of initialization the algorithm finds the correct number of communities for the entire range of values of~$k$ explored.  These results suggest that the underestimation of~$k$ arises not because the correct group assignment fails to maximize the posterior probability~$P(k|A)$, but rather because the Monte Carlo algorithm has not run for long enough to find the maximum.  That is, the method is theoretically sound but the numerical calculation becomes too demanding as $k$ becomes large.  Possibly this problem could be solved with a more efficient Monte Carlo sampling scheme, although it seems likely that some similar issue will eventually arise no matter what sampling scheme is used.  The fundamental problem is that the number of possible group assignments~$k^n$ increases very rapidly with~$k$, so it becomes unfeasible to explore the space of assignments effectively when $k$ is very large.  On the other hand, since, as Figs.~\ref{fig:synthetic}a and~\ref{fig:synthetic}b show, the method only underestimates the value of~$k$ and does not overestimate, the algorithm gives a lower bound on the number of communities in the network, which may well have some utility even when the exact value of $k$ is not found.  Note that the fact that the algorithm underestimates $k$ does not appear to be a result of the initial conditions.  As described in Section~\ref{sec:implementation}, the algorithm starts with an initial number of groups anywhere up to~100, so the initial conditions are at least as likely to overestimate $k$ as underestimate.  It seems likely, therefore, that the underestimates we see in the final results are a consequence of the Monte Carlo sampling method and not of the initial conditions.

The examples in Fig.~\ref{fig:synthetic}a and~\ref{fig:synthetic}b assume so-called assortative network structure, meaning that there are more in-group edges than between-group edges in the network.  Our method, however, should in principle be just as good at finding the number of communities in disassortative cases, where there are more between-group edges, or in mixed assortative/disassortative cases.  Figure~\ref{fig:synthetic}c shows results from tests on networks of mixed type, generated again using a stochastic block model but now taking a diagonally dominant matrix of edge probabilities and permuting the rows and columns to move some of the large matrix entries off the diagonal.  As the figure shows, the algorithm does indeed perform well in this case, indeed it appears to perform better than in the purely assortative case of Figs.~\ref{fig:synthetic}a and~\ref{fig:synthetic}b.  A possible explanation is that in the purely assortative case one can (erroneously) join together groups and produce another assortative network with strong community structure, but in a disassortative or mixed case joining groups is not guaranteed to produce another network with strong structure.

\begin{figure}
\begin{center}
\includegraphics[width=\columnwidth]{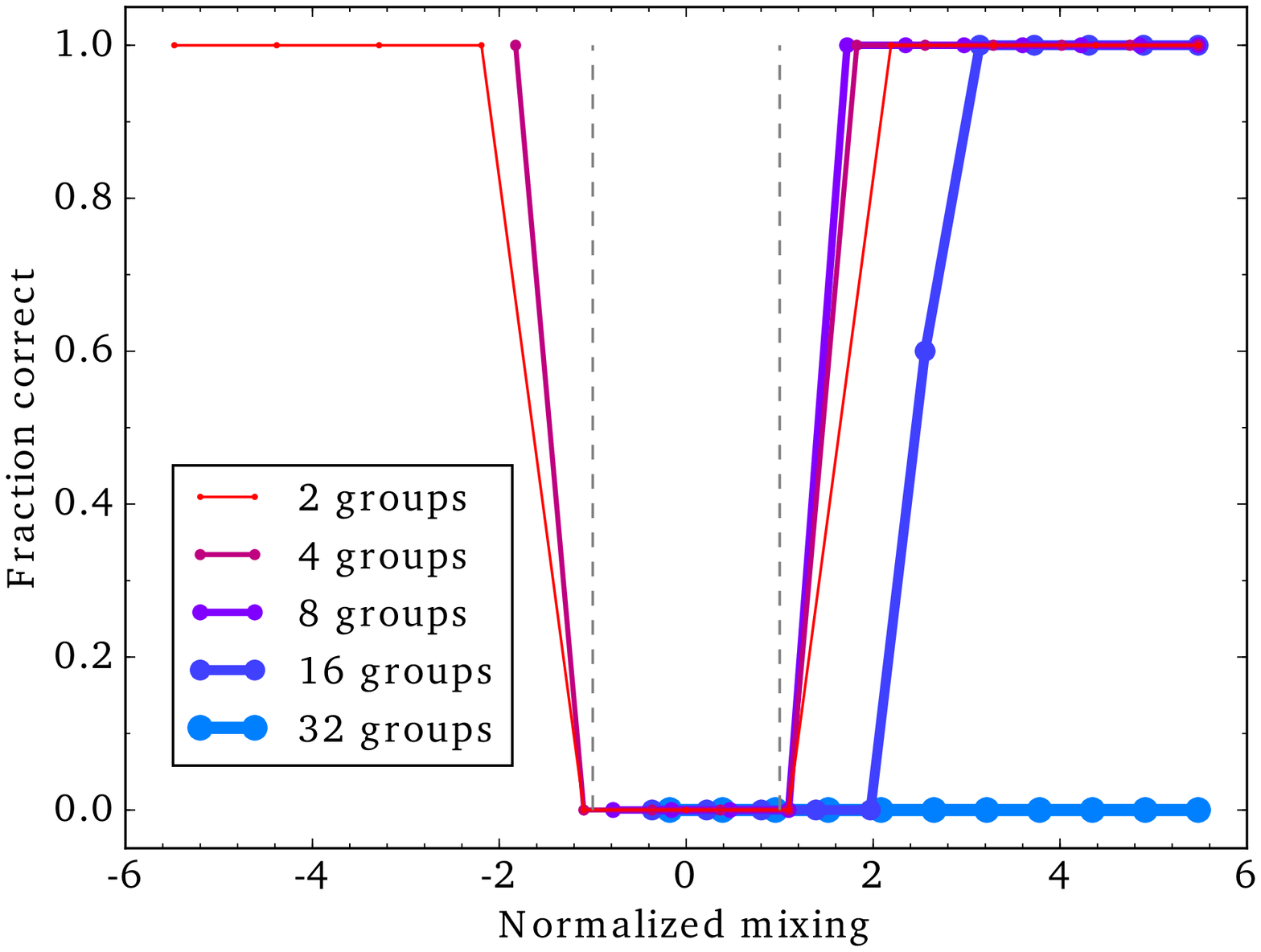}
\end{center}
\caption{Fraction of runs on which the algorithm correctly estimates the number of groups~$k$ in tests on networks generated using the stochastic block model, as a function of the normalized mixing parameter $(\cin-\cout)/\sqrt{c}$.  Networks have $n=1000$ nodes, average degree $c=30$ and $k=2$, 4, 8, 16, 32, with mixing varying from perfect assortativity (right-hand size of the plot) to perfect disassortativity (left-hand side).  The dashed gray lines at $\pm$1 denote the theoretical detectability thresholds.  Fifty Monte Carlo runs were performed for each network of 2000 steps per node each, with the distribution over $k$ calculate from the final 1000 only.}
\label{fig:mixing}
\end{figure}

The parameter values used in each of these examples mean that the networks generated have quite prominent community structure---the number of communities varies from network to network but all have strong structure that should be relatively straightforward to detect.  It is interesting to ask how the method fares if we make the structure weaker.  Figure~\ref{fig:mixing} shows the results of a set of tests on networks with varying difference between the number of in-group and between-group connections.  The horizontal axis is normalized to place the so-called detectability threshold at~$\pm1$.  This is the point at which communities become formally undetectable in the network because the structure is too weak~\cite{DKMZ11a,Massoulie14,MNS15}.  As the figure shows, our method gets the value of $k$ correct virtually all of the time outside of the undetectable region (marked by the vertical dashed lines in the figure), except again for assortative networks with very large numbers of communities (such as the right-hand portion of the curve for $k=32$).

Widely used though it is, one could argue that the stochastic block model is not a very realistic model.  The networks it generates have Poisson degree distributions within each community, for instance, and in the cases studied here we have also limited ourselves to communities of uniform size.  An alternative model that avoids these limitations is the LFR benchmark model of Lancichinetti, Fortunato, and Radicchi~\cite{LFR08}.  This model is essentially a special case of the degree-corrected stochastic block model of Section~\ref{sec:dcsbm}, with both the degrees and the sizes of the communities drawn from power-law distributions, giving the networks similar features to those seen in many real-world examples.

Figure~\ref{fig:synthetic}d shows the results of tests of our method on LFR networks generated with the same model parameters as those used by Lancichinetti and Fortunato~\cite{LF09} in a widely cited study.  The results are similar to those for the stochastic block model: the method performs well for smaller values of the number of groups~$k$ but tends to underestimate as the value of~$k$ gets larger.  If, however, the Monte Carlo algorithm starts with an initial group assignment equal to the planted structure, then it reliably finds the correct number of groups for all values of~$k$, again suggesting that the problem is in the time available for equilibration and sampling, rather than any fundamental issue with the approach.  If one were able to sample from the entire distribution~$P(g,k|A)$ in reasonable time, one should find the correct number of groups.

\subsection{Real-world networks}
\begin{figure*}
\begin{center}
{\hspace{5em}Karate club\hfill American football\hfill Les Mis\'erables\hfill
Word adjacency\hspace{1.2em}\null}\\
\ \\
\includegraphics[width=4cm]{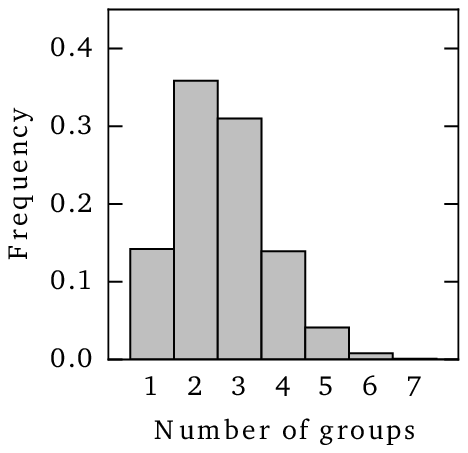}\hfill
\includegraphics[width=4cm]{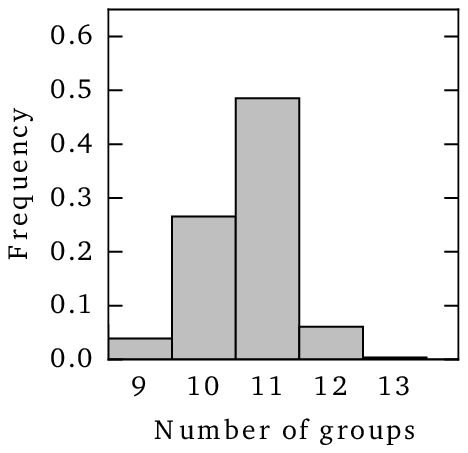}\hfill
\includegraphics[width=4cm]{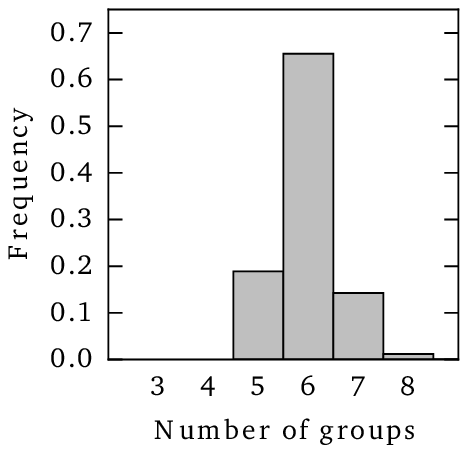}\hfill
\includegraphics[width=4cm]{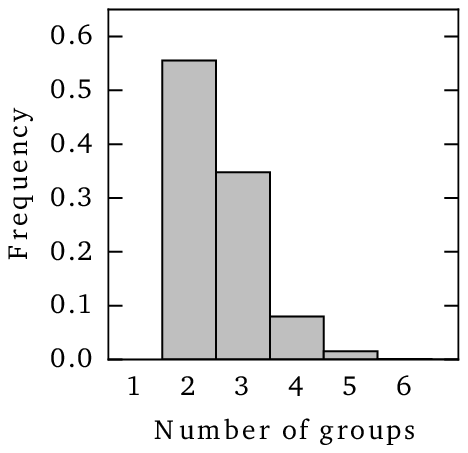}\\
\ \\
\includegraphics[width=4cm]{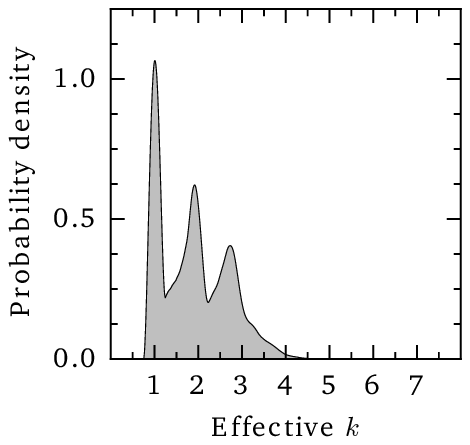}\hfill
\includegraphics[width=4cm]{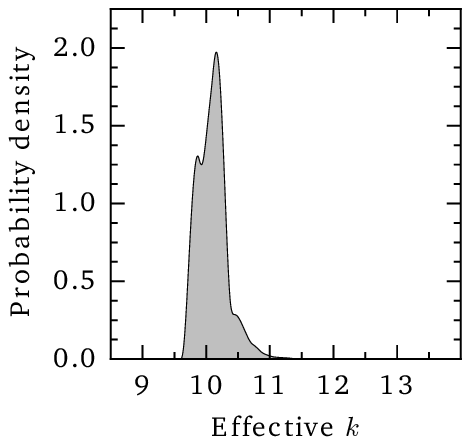}\hfill
\includegraphics[width=4cm]{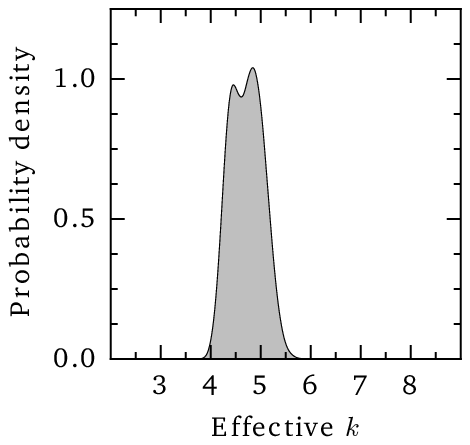}\hfill
\includegraphics[width=4cm]{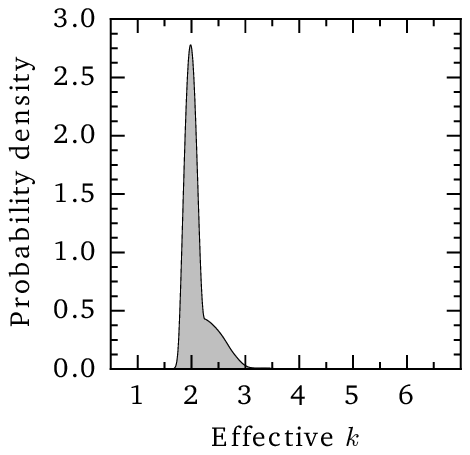}\\
\ \\
\ \\
\includegraphics[width=3.5cm]{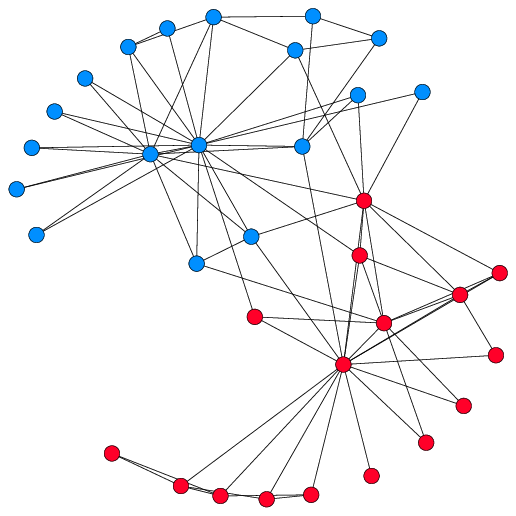}\hfill
\includegraphics[width=3.5cm]{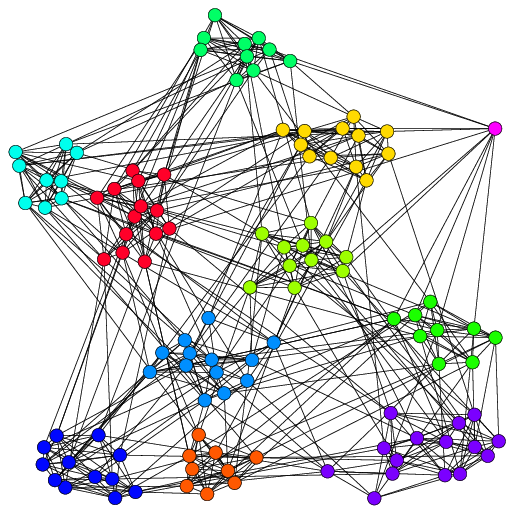}\hfill
\includegraphics[width=3.5cm]{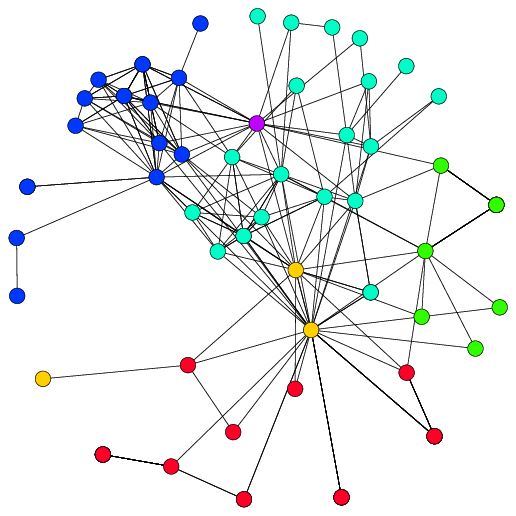}\hfill
\includegraphics[width=3.5cm]{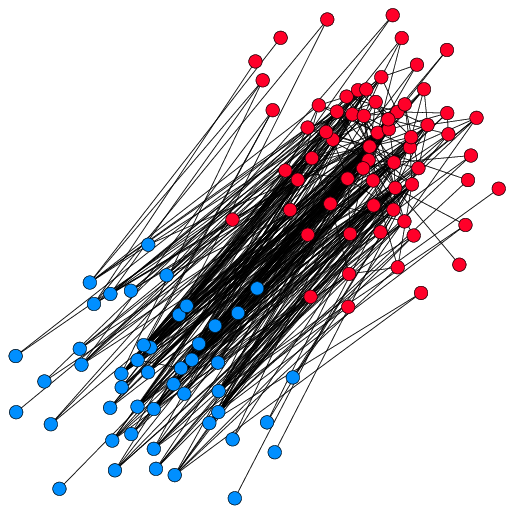}\\
\ \\
\includegraphics[height=4cm]{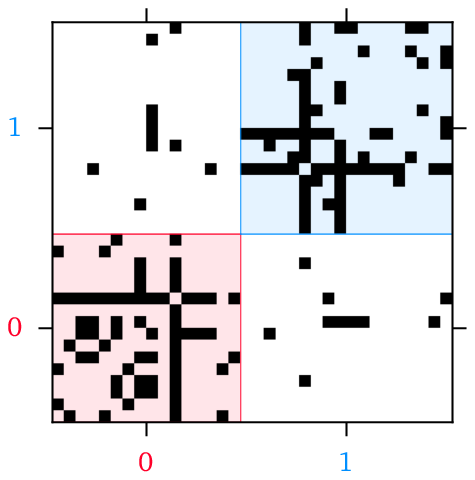}\hfill
\includegraphics[height=4cm]{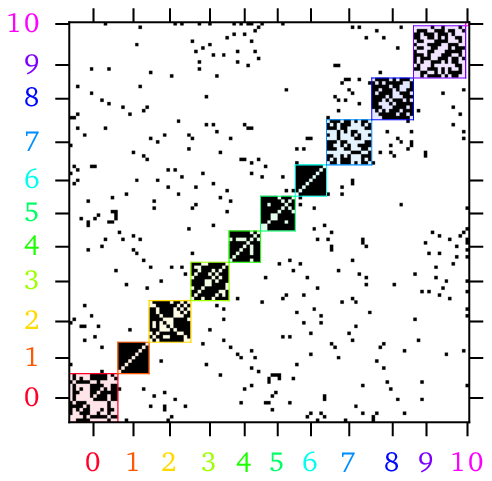}\hfill
\includegraphics[height=4cm]{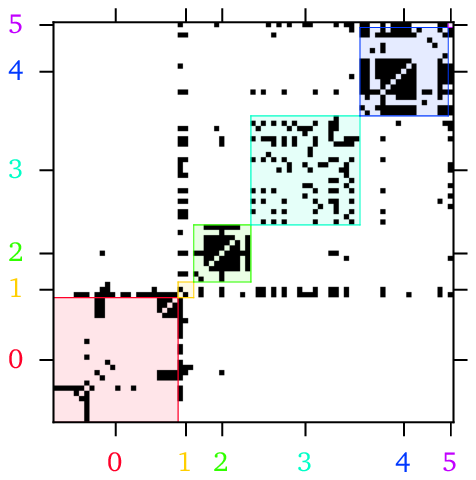}\hfill
\includegraphics[height=4cm]{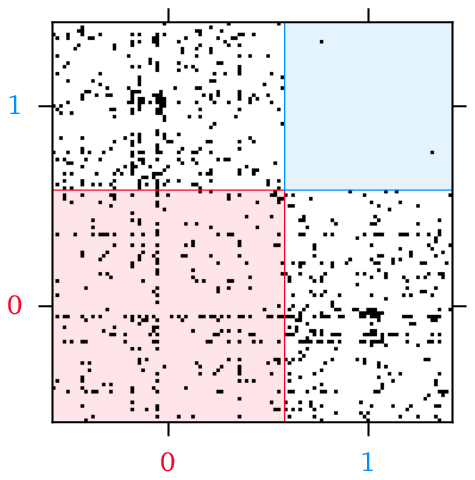}\\
\ \\
\ \\
\qquad\includegraphics[width=2.5cm]{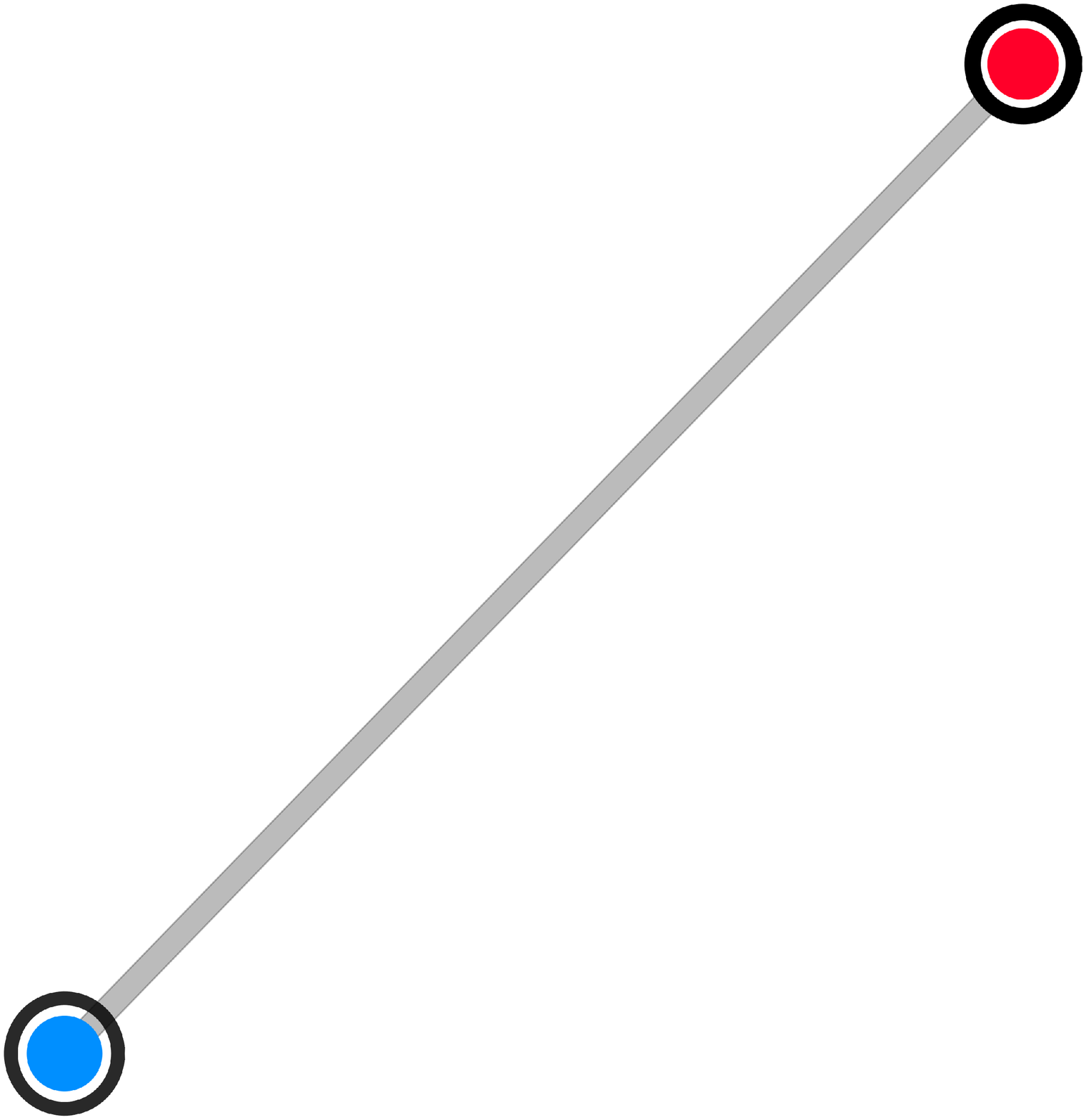}\hfill
\includegraphics[width=2.5cm]{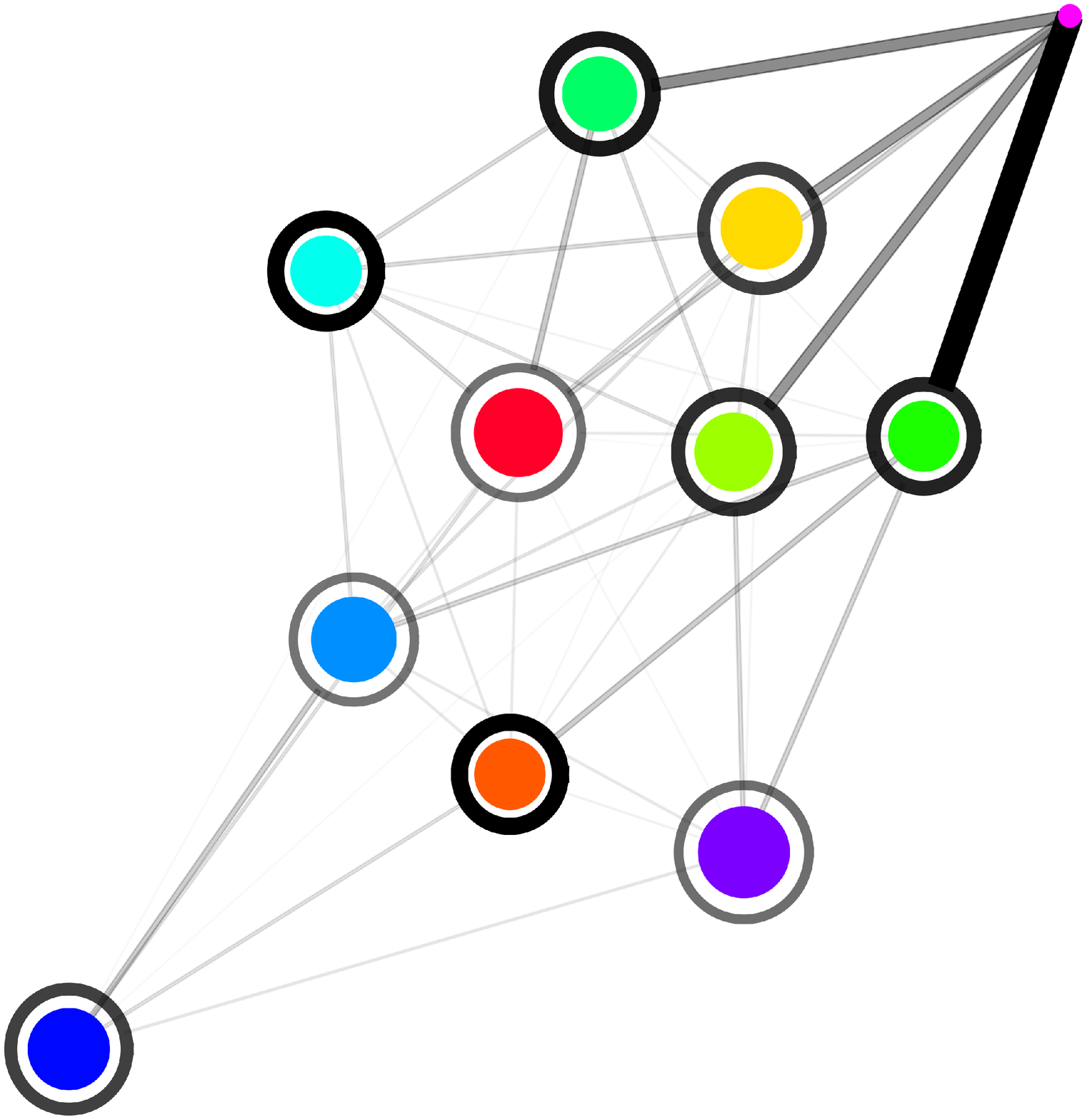}\hfill
\includegraphics[width=2.5cm]{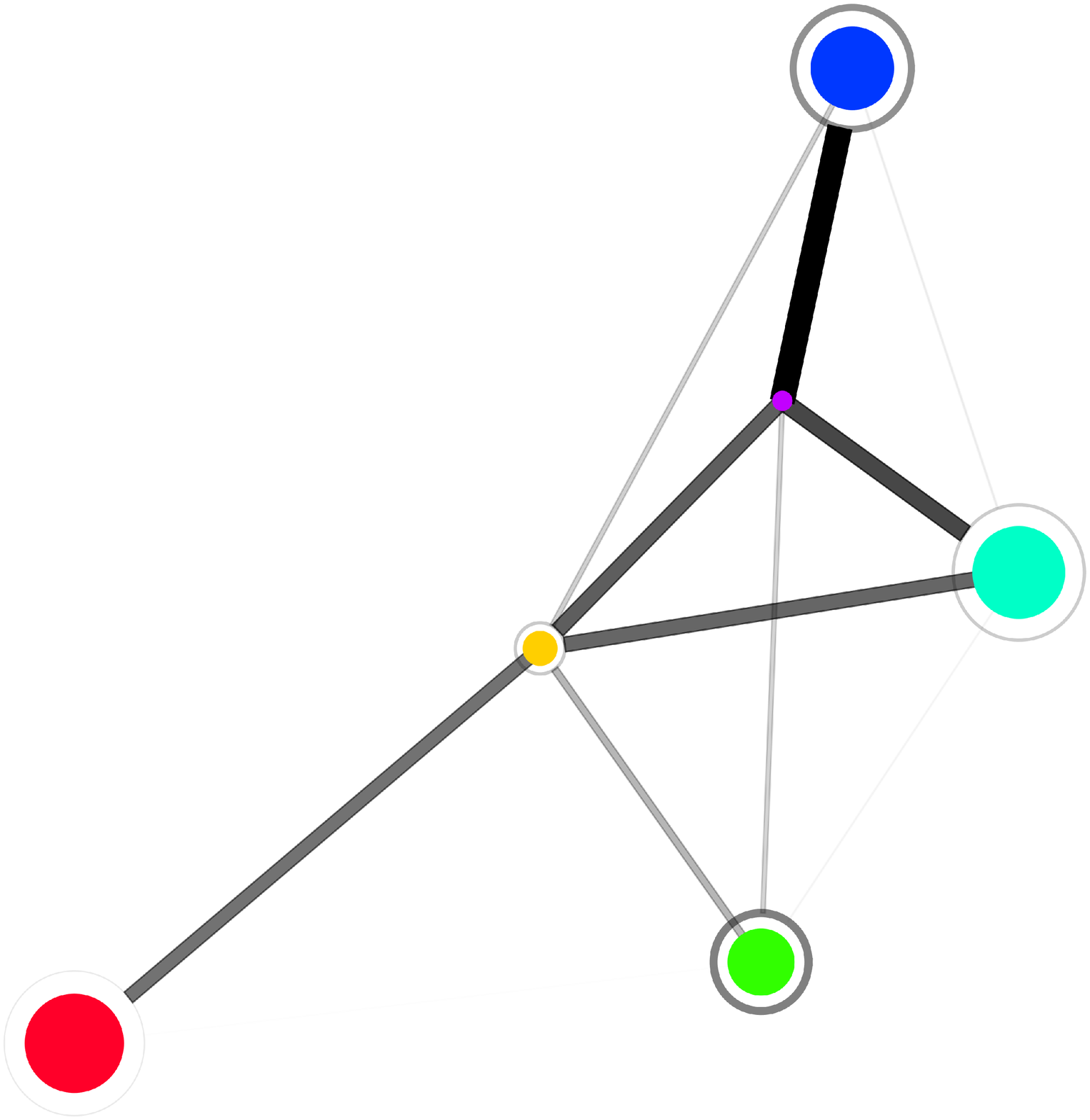}\hfill
\includegraphics[width=2.5cm]{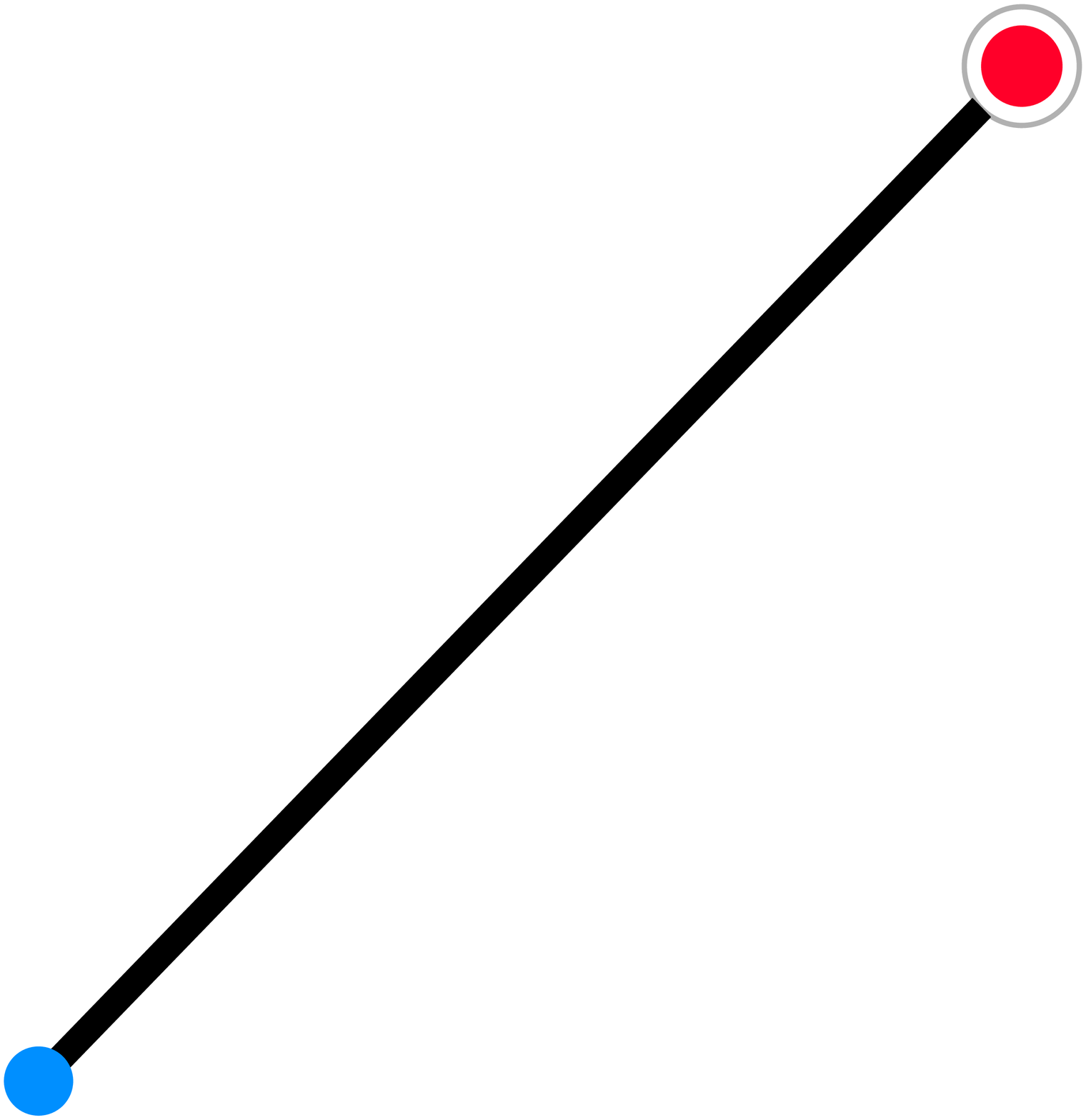}\qquad\null
\end{center}
\caption{Results for four real-world networks.  Each column shows results for one network, as indicated.  From top to bottom the results shown are the posterior probability distribution of the number of groups, the distribution of the effective number of groups calculated using the entropy measure of Eq.~\eqref{eq:keff}, the maximum likelihood community structure, the adjacency matrix, and the ``meta-network'' representation of the communities and their pattern of connectivity.\label{fig:real}}
\end{figure*}

As a complement to the synthetic tests of the previous section, as have also tested our method on a range of real-world networks.  There exist a number of well studied example networks in the literature that have widely agreed upon ground-truth community divisions, based in part on knowledge of the specific systems the networks describe and in part on consensus derived from repeated analyses with many different community detection algorithms.  Figure~\ref{fig:real} shows results for four such networks.

The first column in Fig.~\ref{fig:real} shows results for tests on the ``karate club'' network of Zachary~\cite{Zachary77}, perhaps the best known and most widely used benchmark of community detection.  This small social network is universally agreed to contain two clear communities, and when applied to the network our method firmly favors $k=2$.  The top panel in the figure shows a histogram of the values of $k$ sampled by the Monte Carlo algorithm on this network, and the probability shows a clear peak for two communities.

It could be argued, however, that looking directly at the values of~$k$ generated by the Monte Carlo algorithm has the potential to be misleading in some cases.  Imagine, for instance, that a network breaks apart into two large groups that occupy most of the network, plus a third group with only a very few nodes in it.  In this situation one might be justified in saying that the network really only contains two groups, not three.  One can capture this kind of situation by defining an effective number of groups
\begin{equation}
k_\textrm{eff} = e^S, \qquad
S = - \sum_{r=1}^k {n_r\over n} \log {n_r\over n}.
\label{eq:keff}
\end{equation}
Here $S$ is the entropy of the group assignment, which has a maximum value of $\log k$ when the groups are equally sized, so that $k_\textrm{eff} = e^S = k$ in this case.  On the other hand, if there are a few small groups in the network and the remainder are large and equally sized then the measure will ignore the small groups to a great extent and $k_\textrm{eff}$ will be roughly equal to the number of large groups only.

The second panel in column~1 of Fig.~\ref{fig:real} shows results for this alternative measure of group number, as calculated using our Monte Carlo algorithm on the karate club network.  It is straightforward to show that $k_\textrm{eff} \le k$ strictly, so by necessity the distribution in the second panel is to the left of that in the top panel.  There are three clear peaks visible in the distribution of~$k_\textrm{eff}$, corresponding to states with one, two, and three groups, so it seems reasonable to assume that these are ``real'' divisions of the network.  The peak for $k_\textrm{eff}=1$ is the highest, but the area under the peak for $k_\textrm{eff}=2$ is greater, so two groups, which is the widely accepted number, still seems to be preferred overall.

The third panel in column~1 shows the group assignment~$g$ with the maximum likelihood sampled by the Monte Carlo algorithm, which in this case corresponds closely to the accepted community structure of the karate club network.  The fourth panel shows an alternative visualization of the community structure, a plot of the adjacency matrix of the network where the rows and columns have been ordered so as to put the two groups in contiguous blocks.  As we can see, this places most edges within blocks and only a few between blocks, as we would expect for a network with strong community structure.  Finally, in the bottom panel of the column, we show a visualization of the group structure itself, a ``meta-network'' in which the nodes represent the groups and edges represent connections between groups.  In this simple case the meta-network does not offer much insight, since it consists of just two meta-nodes and a single edge, but in more complicated situations with larger numbers of groups---including some of the others in Fig.~\ref{fig:real}---it can be a useful tool.

The remaining three columns of Fig.~\ref{fig:real} show corresponding analyses for three further networks: the American college football network studied in~\cite{GN02}, the network of fictional character interactions in the novel \textit{Les Mis\'erables} by Victor Hugo~\cite{Knuth93}, and a disassortative example, the network of word adjacencies of adjectives and nouns in the novel \textit{David Copperfield} by Charles Dickens~\cite{Newman06c}.  In each case the algorithm finds the accepted number of communities---six, eleven, and two, respectively.  The distributions of~$k_\textrm{eff}$ largely agree with those for~$k$, though for the \textit{Les Mis\'erables} and college football networks they favor five and ten groups respectively, one less than the peak in the distribution of $k$ in each case, perhaps suggesting that these networks have one group that is small enough to be neglected.

\begin{figure}
\begin{center}
\includegraphics[width=\columnwidth]{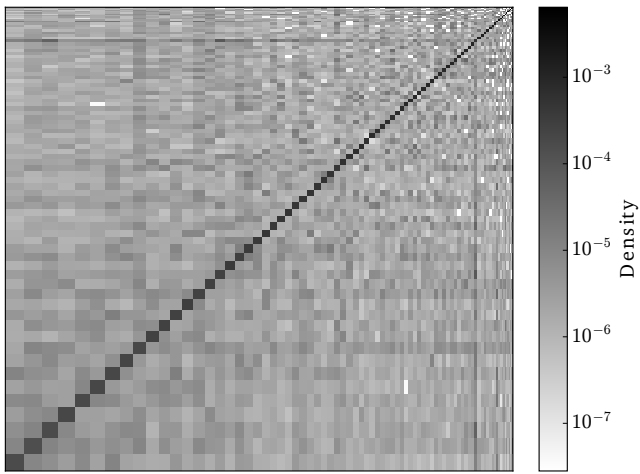}
\end{center}
\caption{A plot of the matrix~$\omega_{g_ig_j}$ of connection parameters for a large network of copurchased products on Amazon.com.  In this network nodes represents products and edges join pairs of products frequently purchased by the same buyer.  In this calculation we performed ten runs of $10\,000$ Monte Carlo steps per node each and we display results from the run with the highest average likelihood during the last 1000.  The calculation found 81 groups in total.  Color is on a log scale for clarity.}
\label{fig:amazon}
\end{figure}

These examples are all relatively small networks, up to around a hundred nodes in the larger cases, but our method is applicable in principle to much larger networks.  As an example, Fig.~\ref{fig:amazon} shows results for a network of e-commerce data, a copurchasing network of items sold by the online retailer Amazon.com~\cite{YL15}.  In this network, which comes from the Stanford Large Network Dataset collection, the nodes represent $334\,863$ products for sale on the Amazon web site and edges between them indicate products that were frequently purchased by the same buyer.  The figure shows a visualization of the inferred values of the edge probability parameters~$\omega_{g_ig_j}$, again with the columns ordered so as to make the groups contiguous.  As we can see, there appears to be strong assortative structure in the network, with the algorithm finding 81 groups in this case.

\section{Conclusions}
In this paper we have described a method for determining the number of communities in a network with community structure.  The method relies on a combination of Bayesian inference applied to the degree-corrected stochastic block model and a novel Monte Carlo algorithm.  Much of the method's success turns on the appropriate choice of prior probability for the number of groups and we describe a variation on the ``restaurant'' processes of traditional model selection that appears to work well.  We have illustrated the performance of the method with applications to a wide range of networks, including a diverse set of synthetic test networks and a number of real-world examples, one with over $300\,000$ nodes.

The primary limitation of the method as described is that the Monte Carlo algorithm appears not to equilibrate fully when the number of groups becomes large.  A possible objective for future work, therefore, would be to find a method or algorithm that could sample the posterior distribution over group assignments faster, which would allow us to better apply the method to networks with large numbers of groups.

\begin{acknowledgments}
The authors thank Tiago Peixoto and Jia-Rong Xie for helpful conversations.  This work was funded in part by the US National Science Foundation under grants DMS--1107796 and DMS--1407207 (MEJN), the UK Engineering and Physical Sciences Research Council under grant EP/K032402/1 (GR), the James S. McDonnell Foundation (MAR), the Simons Foundation (MEJN), and the Advanced Studies Centre at Keble College, Oxford (MEJN and GR).
\end{acknowledgments}

\end{document}